\documentclass[sigconf,nonacm]{acmart}

\AtBeginDocument{%
  \providecommand\BibTeX{{%
    \normalfont B\kern-0.5em{\scshape i\kern-0.25em b}\kern-0.8em\TeX}}}

\setcopyright{acmlicensed}
\copyrightyear{2018}
\acmYear{2018}
\acmDOI{XXXXXXX.XXXXXXX}

\acmConference[ICSE 2024]{46th International Conference on Software Engineering}{April 2024}{Lisbon, Portugal}
\acmBooktitle{2nd International Workshop on Responsible AI Engineering, ICSE 2024,
 April 2024, Lisbon} 
\acmISBN{978-1-4503-XXXX-X/18/06}
\usepackage{subcaption}
\usepackage{graphicx} 
\usepackage{longtable}
\usepackage{adjustbox}
\begin{document}

\title{Responsible AI Governance: A Systematic Literature Review}

% \affiliation{%

\author{Amna Batool}
\affiliation{%
  \institution{CSIRO's Data61}
  \streetaddress{}
  \city{Melbourne}
  \country{Australia}}
\email{amna.batool@data61.csiro.au}

\author{Didar Zowghi}
\affiliation{%
  \institution{CSIRO's Data61}
  \streetaddress{}
  \city{Melbourne}
  \country{Australia}}
\email{didar.zowghi@data61.csiro.au}

\author{Muneera Bano}
\affiliation{%
  \institution{CSIRO's Data61}
  \streetaddress{}
  \city{Melbourne}
  \country{Australia}}
\email{muneera.bano@data61.csiro.au}

%% \renewcommand{\shortauthors}{Trovato and Tobin, et al.}

\begin{abstract}
 As artificial intelligence transforms a wide range of sectors and drives innovation, it also introduces complex challenges concerning ethics, transparency, bias, and fairness. The imperative for integrating Responsible AI (RAI) principles within governance frameworks is paramount to mitigate these emerging risks. While there are many solutions for AI governance, significant questions remain about their effectiveness in practice. Addressing this knowledge gap, this paper aims to examine the existing literature on AI Governance. The focus of this study is to analyse the literature to answer key questions: WHO is accountable for AI systems' governance, WHAT elements are being governed, WHEN governance occurs within the AI development life cycle, and HOW it is executed through various mechanisms like frameworks, tools, standards, policies, or models. Employing a systematic literature review methodology, a rigorous search and selection process has been employed. This effort resulted in the identification of 61 relevant articles on the subject of AI Governance. Out of the 61 studies analysed, only 5 provided complete responses to all questions. The findings from this review aid research in formulating more holistic and comprehensive Responsible AI (RAI) governance frameworks. This study highlights important role of AI governance on various levels specially organisational in establishing effective and responsible AI practices. The findings of this study provides a foundational basis for future research and development of comprehensive governance models that align with RAI principles.

\end{abstract}

\begin{CCSXML}
<ccs2012>
 <concept>
  <concept_id>00000000.0000000.0000000</concept_id>
  <concept_desc>Do Not Use This Code, Generate the Correct Terms for Your Paper</concept_desc>
  <concept_significance>500</concept_significance>
 </concept>
 <concept>
  <concept_id>00000000.00000000.00000000</concept_id>
  <concept_desc>Do Not Use This Code, Generate the Correct Terms for Your Paper</concept_desc>
  <concept_significance>300</concept_significance>
 </concept>
 <concept>
  <concept_id>00000000.00000000.00000000</concept_id>
  <concept_desc>Do Not Use This Code, Generate the Correct Terms for Your Paper</concept_desc>
  <concept_significance>100</concept_significance>
 </concept>
 <concept>
  <concept_id>00000000.00000000.00000000</concept_id>
  <concept_desc>Do Not Use This Code, Generate the Correct Terms for Your Paper</concept_desc>
  <concept_significance>100</concept_significance>
 </concept>
</ccs2012>
\end{CCSXML}

% \ccsdesc{Social and professional topics ~ User characteristics}
% \ccsdesc{Software and its engineering ~ Software creation and management ~ Designing software ~ Requirements analysis}
% \ccsdesc{Computing methodologies ~ Artificial intelligence}

\keywords{Artificial Intelligence, AI Governance, responsible AI}



\maketitle

\section{Introduction}
Artificial intelligence (AI) has emerged as one of the most important technologies in many businesses and has grown to be an integral part of our society \cite{lu2022responsible}. However, the risks and negative effects of AI are growing with its widespread application in a variety of sectors, such as  autonomous cars \cite{lutge2021ai4people}, healthcare \cite{reddy2020governance}, finance \cite{lee2020access}, and other areas. There are different repositories\footnote{https://incidentdatabase.ai} \footnote{https://www.aiaaic.org/aiaaic-repository} on AI incident databases that contains over 2000 AI incidents. \\
Responsible AI refers to developing, deploying, and utilizing AI in order to ensure it is developed and used in a manner that aligns with human values, societal norms, legal standards, and mitigates potential risks and negative consequences associated with it, such as bias, discrimination, and lack of transparency \cite{lu2022responsible}, \cite{nationalai2023}, \cite{zowghi2023diversity}. In order to deal with the AI risks, ethical use of AI has become an essential demand \cite{hagendorff2020ethics}. As a result, governmental and international organisations, including the European Commission \footnote{https://digital-strategy.ec.europa.eu/en/library/ethics-guidelines-trustworthy-ai}, the Organisation for Economic Co-operation and Development [OECD] \footnote{https://oecd.ai/en/ai-principles}, Australian government \footnote{https://www.industry.gov.au/publications/australias-artificial-intelligence-ethics-framework/australias-ai-ethics-principles}, professional bodies (the Institute of Electrical and Electronics Engineers [IEEE]), and various companies have published their ethical AI principles and guidelines \cite{fjeld2020principled}, \cite{jobin2019global}.

The effort goes beyond just technical concerns and includes the development of strong AI governance solutions and techniques, which are considered essential for directing responsible use of AI technology, striking a balance between innovation and ethical behaviour, and avoiding unforeseen effects \cite{mantymaki2022putting}. Ethics principles and guidelines must be embedded in AI systems through proper AI governance techniques \cite{cath2018governing}. AI governance can be understood as encompassing a set of regulations, methods, procedures, and technological mechanisms utilized to guarantee that an organisation's utilization of AI technologies is consistent with its strategies, goals, and principles \cite{mantymaki2022putting}. This encompasses adhering to legal requirements and upholding the ethical AI principles adopted by the organisation \cite{mantymaki2022putting}. There have been numerous research studies done on proposing different AI governance solutions to govern AI systems  \cite{koniakou2023rush}, including the ECCOLA method for the governance of ethical and trustworthy AI systems \cite{agbese2023governance}, \cite{agbese2021governance}, the framework for AI governance proposed for AI safety domains \cite{maas2019innovation}, a governance model for the AI application in healthcare sectors \cite{reddy2020governance}. However, there are multiple challenges, including transparency, explainability, and accountability of algorithms in AI governance, that need to be addressed \cite{larsson2020transparency}, \cite{kroll2021outlining}. In addition to these challenges, AI governance also faces diversity and inclusion issues such as lack of fairness, poor reliability, lack of inclusive solutions, and so on \cite{carter2020regulation}. Governance is recognised to be the least focused area in AI Diversity and Inclusion \cite{shams2023ai}.

To study the many facets of this developing discipline, this research paper dives into the topic of responsible AI governance. The establishment of structures and procedures at various levels—team, organisation, and industry, to make sure the development and implementation of AI systems adhere to ethical principles in line with AI governance patterns for responsible AI can be referred to as responsible AI governance \cite{lu2022responsible}. An approach referred to as 3W1H (Who, What, When, and How) has been adopted in this study to analyse the existing AI governance solutions from the literature. In this approach, 3W covers three questions: who is governing (stakeholders), what should be governed (humans, data, system, or process), as proposed by Zowghi and da Rimini \cite{zowghi2023diversity}, and when is it being governed (i.e. at what stage of AI development life cycle)? 1H covers: How is AI being governed (i.e. frameworks, standards, regulations, or policies, etc. A comprehensive analysis of literature using 3W1H approach is presented in this paper, and the categorisation of key elements from 3W1H questions is covered under different layers of governance from the study by Lu et al. \cite{lu2022responsible}: team-level, organisation-level, industry-level, national-level, and international-level, which are explained later in this paper.

This study aims to summarize and synthesize current AI governance solutions (i.e. frameworks, tools, models, and policies), examine challenges in existing AI governance solutions, and offer insights based on the answers to the 3W1H questions. The main contributions of this study are: (1) A comprehensive analysis of 61 research papers selected from the academic literature has been presented, (2) The challenges or limitations of existing AI governance solutions are also presented, and (3) The categorisation of key elements from 3W1H questions are presented under five levels of governance.

% \begin{itemize}
%     \item A comprehensive analysis of 61 research studies selected from the academic literature has been performed and presented in this paper.
%     \item After performing analysis of 3W1H questions, the categorisation of key elements from 3W1H questions (who is governing and how is it governed) has been done under five layers of governance patterns for responsible AI.
%     \item The challenges or limitations of existing AI governance solutions (frameworks, models, tools, policies) have been observed and added in this study with useful insights on future AI governance research directions. 
% \end{itemize}

This paper is organised as follows: Section two presents the research methodology along with research questions and data extraction. Section three presents the data analysis of the 3W1H approach carried out on 61 selected studies and the categorisation of AI governance solutions under five AI governance levels. The research results, where the answers to the research questions are added, are in Section 4. Section five concludes the study with future work.

\section{Research Methodology}
This section presents the methodology followed in conducting this systematic literature review (SLR) following the guidelines established by Kitchenham et al. \cite{kitchenham2009systematic}. Figure~\ref{fig:one} presents the methodology steps followed and is explained below: 

\begin{figure}
\centering
  \includegraphics[width=0.46\textwidth]{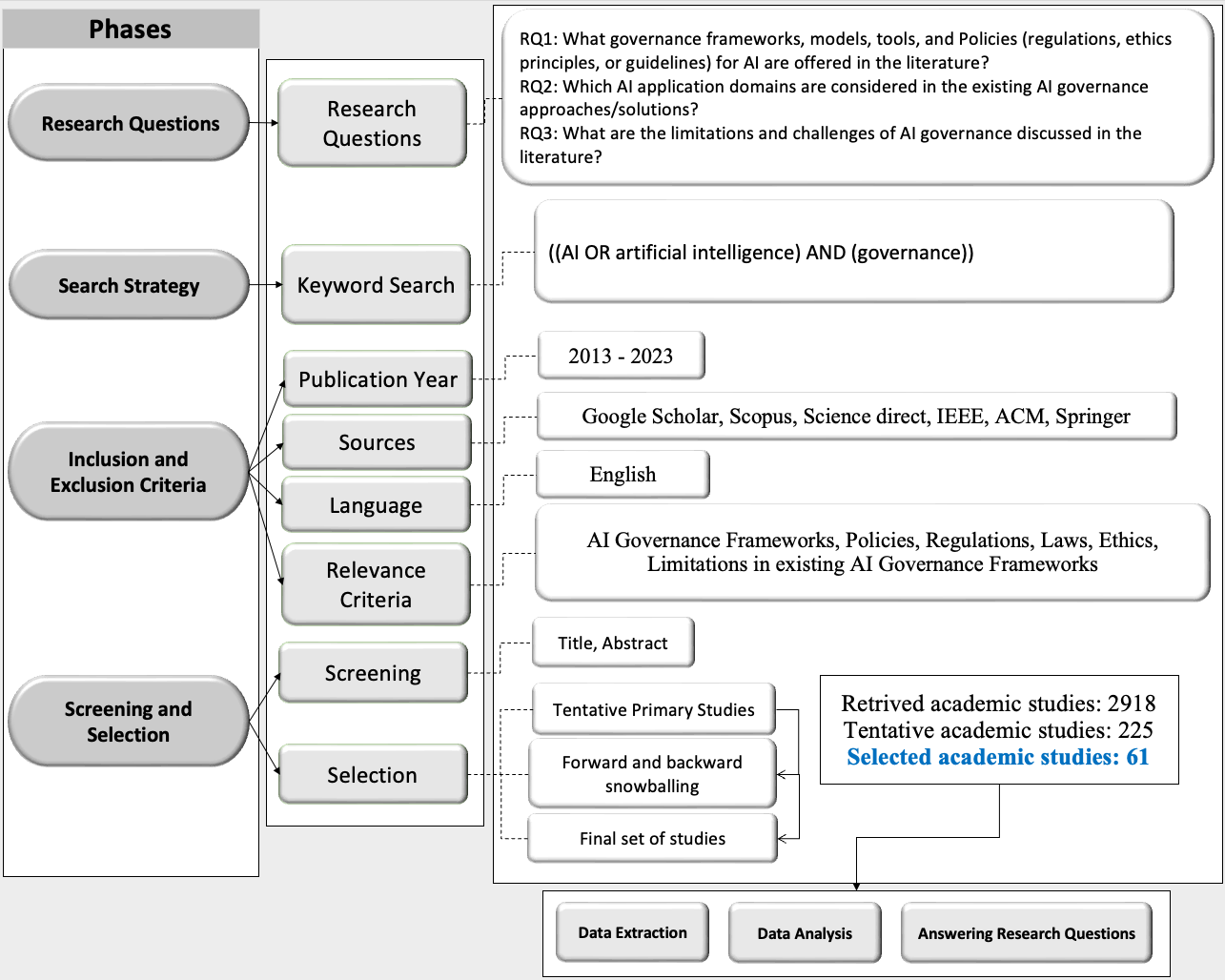}
  \caption{Methodology Overview}
      \label{fig:one}
\end{figure}

\textbf{Research Questions}: The following research questions have guided this systematic review: 

\begin{itemize}
    \item RQ1: What governance frameworks, models, tools, and Policies (regulations, ethics principles, or guidelines etc) for AI are offered in the literature?
    \item RQ2: Which target AI application domains are considered in the existing AI governance approaches/solutions?
    \item RQ3: What are the limitations and challenges of AI governance discussed in the literature?
\end{itemize}

\textbf{Search Strategy}: The keywords used to search for the articles in search engines and digital libraries are: 
((AI OR artificial intelligence) AND (governance)) 

\textbf{Inclusion and Exclusion Criteria}: A wide range of search engines and digital libraries have been used to identify the relevant articles. Then, the inclusion criteria of year of publication and relevance were applied. The articles published in English between 2013 to 2023 in Google Scholar, Scopus, Science Direct, IEEE, and ACM were considered while determining inclusion criteria. Moreover, the papers relevant to the research focus were considered for inclusion, while book chapters, reports, and documentary articles were excluded. Papers that meet the relevance criteria were  included in this study. All other papers that did not fit these requirements were omitted. Consequently, publications focusing on such topics as environmental governance, climate change, ecology, etc., were excluded. 

\textbf{Screening and Selection}: Guidelines for Systematic Literature Review established by Kitchenham et al. \cite{kitchenham2009systematic} mainly focus on empirical and primary studies. Considering the limited number of empirical and primary studies on AI governance, the research scope was extended to include non-empirical and secondary studies. This approach enabled the coverage of a broader spectrum of knowledge on the topic. To aid clarity and provide insightful analysis, this systematic review has distinctly categorised and presented these varying types of research for the readers. The screening process resulted in 2918 academic papers, and the selection process ended with the selection of 225 articles. Out of 225 resulting articles, 61 studies were selected (both Empirical - E and Non-Empirical - N) for this literature study after applying forward and backward snowballing and are added to this link \footnote{https://github.com/amnacsiro/SLR-AI-Governance/tree/main}. 

\begin{figure}
\centering
  \includegraphics[width=0.4\textwidth]{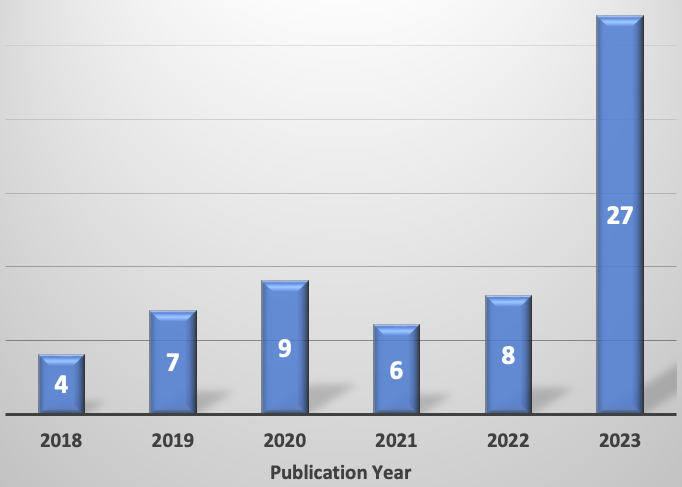}
  \caption{Publication Year of Selected Set of Studies}
      \label{fig:two}
\end{figure}

\begin{figure}
\centering
  \includegraphics[width=0.4\textwidth]{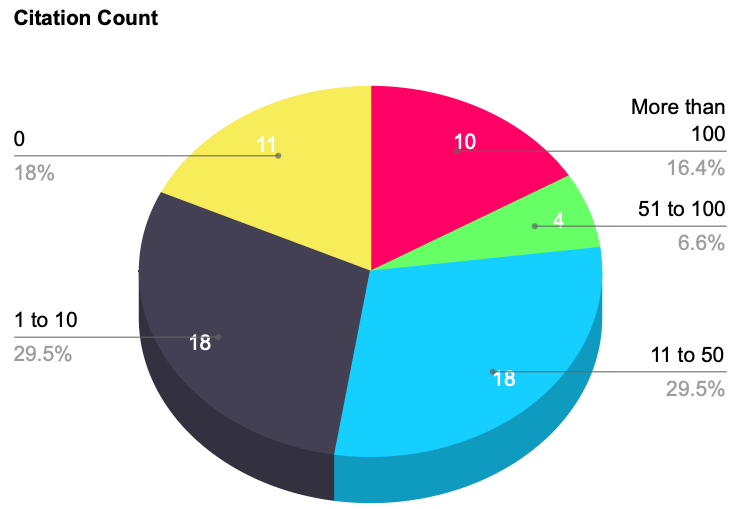}
  \caption{Citations}
      \label{fig:three}
\end{figure}

\begin{figure*}%
    \centering
    \subfloat[\centering Empirical Studies]{{\includegraphics[width=7cm]{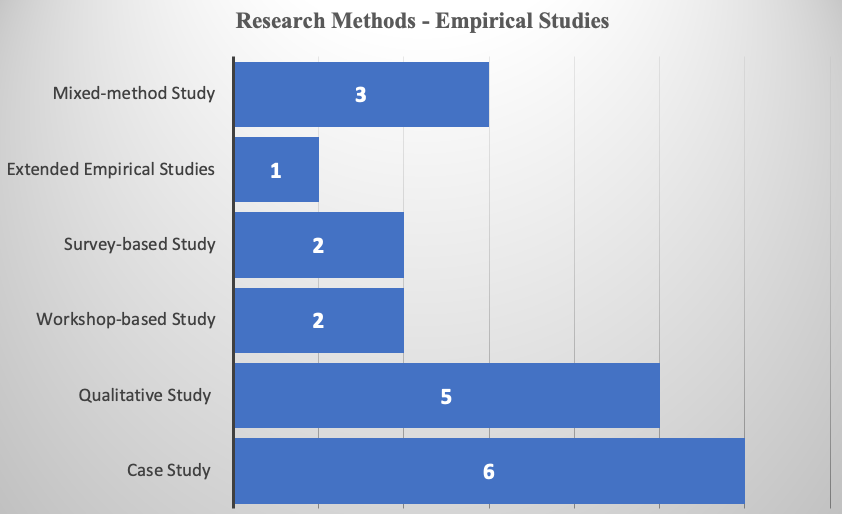} }}%
    \subfloat[\centering Non-Empirical Studies]{{\includegraphics[width=7cm, height=4.3cm]{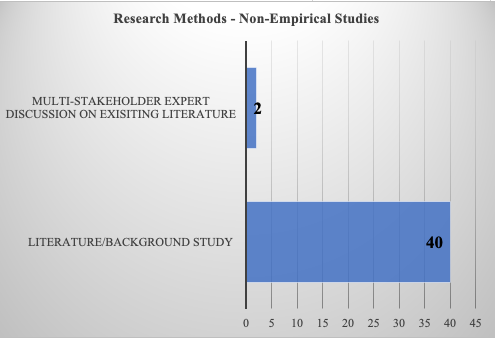}}}%
    \caption{Research Methods: Empirical Studies (a) Non-Empirical Studies (b)}%
    \label{fig:four}%
\end{figure*}

\textbf{Data Extraction}
First, the year of publication of the selected collection of studies is determined. The studies' publication year is shown in Figure~\ref{fig:two}, which indicates that, out of the 61 papers, the greatest number of studies (27) were published in 2023, and the lowest number (4) in 2018. It has been noted that since 2018, AI governance has received much attention and is an increasingly prevalent topic today. Regarding citations, Figure~\ref{fig:three} shows that 4 studies obtained 51 to 100 citations, while 10 papers received more than 100 citations throughout the past ten years covered for this SLR. As the selected set of 61 studies contains both empirical (E) and non-empirical (N) research \footnote{https://github.com/amnacsiro/SLR-AI-Governance/tree/main}. Figure~\ref{fig:four} presents the number of empirical studies 19, including case study papers, qualitative studies, survey and workshop-based studies, etc., and non-empirical studies 42, which are literature/background studies and discussions on existing literature on AI governance.

\section{Data Analysis}
This section presents the data analysis of 61 selected studies through the 3W1H approach and also presents the categorisation of stakeholders based on "Who is governing?" and artefacts based on "How is it governed?" under five levels of AI governance.

\subsection{3W1H Approach:}
The 3W1H approach consists of questions, Who, What, When, and How, which have been explored for the selected set of 61 studies using thematic analysis. This analysis involved a comprehensive examination of the literature to identify recurring themes related to AI governance. "Who" focuses on the stakeholders (key AI oversight roles) who should be involved in governing to oversee if AI complies with standards or regulations for responsible AI. These stakeholders include responsible research and innovation officers, data management scientists, AI governance committee members etc. Next, "What" covers the four pillars of the AI ecosystem, which are humans, data, processes, and systems. The selection of these four pillars is grounded in a thorough review of the existing literature and a consideration of the fundamental elements that constitute the AI ecosystem \cite{zowghi2023diversity}. Aligning with Zowghi and da Rimini's study \cite{zowghi2023diversity}, each pillar is further explained below. 

\begin{itemize}
    \item Humans: Acknowledging the centrality of human involvement in AI development and deployment, this pillar encompasses stakeholders (operational AI practitioners) involved in the day-to-day technical implementation and operational aspects of AI projects, such as AI developers, AI designers, etc. From the previous question, "Who is governing?", which includes key AI oversight roles, this pillar is linked with a focus on observing "Who is governing whom?", which means whether key AI oversight roles are governing operational AI practitioners for ensuring that AI systems they are developing adhere to ethical standards and regulatory requirements.  
    \item Data: This pillar emphasizes the importance of responsible data acquisition, processing, and utilization. Effective AI governance requires a focus on data quality, privacy, and security to mitigate biases and promote fair and transparent AI decision-making.
    \item Process: The AI practices involved in AI development and deployment that are overseen by stakeholders to guarantee they comply with ethical standards, legal requirements, and organisational goals are included in the 'Process' pillar. For instance, if the stakeholders involved in the preceding question on "Who is governing?", are taking responsibility for governing the pre-development, during, and post-development processes of AI systems development.
    \item System: This pillar encompasses the AI systems, which should be governed in consideration of their robustness, interpretability, and accountability to guarantee their responsible and reliable operation.
\end{itemize}

Next, "When" represents the four development stages of AI to which the particular governance solutions should be applied. The four stages of AI development are: the initial stage (planning, approval, and data collection), design, development (build models, verify, and validate), and deployment (monitor, and use), in alignment with established frameworks and guidelines from organisations such as NIST \cite{NIST_AIRMF}, OECD \cite{OECD_AI_Society}. Lastly, the "How" explores the artefacts of governance solutions for AI , such as policies, ethics principles, tools or frameworks, standards, laws, regulations, models, etc. 

% \begin{figure*}%
%     \centering
%     \subfloat[\centering 3W: Who, What, When]{{\includegraphics[width=5cm, height=6.2cm]{3W-LASTTTTTTTTTTTTTTTTT.png} }}%
%     \subfloat[\centering 1H: How]{{\includegraphics[width=12.5cm]{1H-LASTTTTTTTTTTTTTTTTT.png} }}%
%     \caption{3W1H Analysis: Who, What, When (a:left), and How (b:right)}%
%      \label{fig:six}%
% \end{figure*}

\begin{figure*}
\centering
  \includegraphics[width=0.8\textwidth]{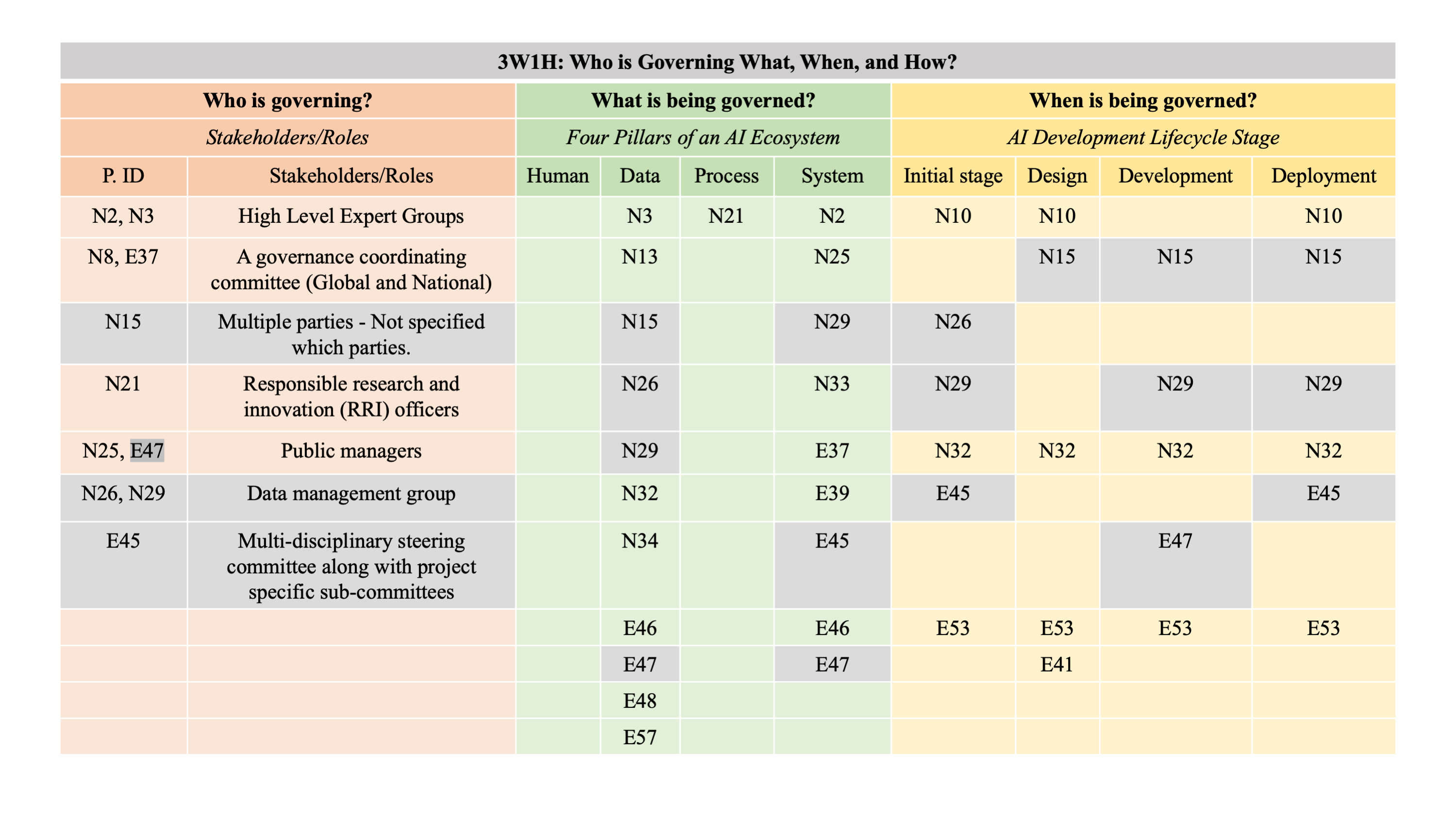}
  \caption{3W1H Analysis: Who, What, and When}
      \label{fig:six}
\end{figure*}

\begin{figure*}
\centering
  \includegraphics[width=0.8\textwidth]{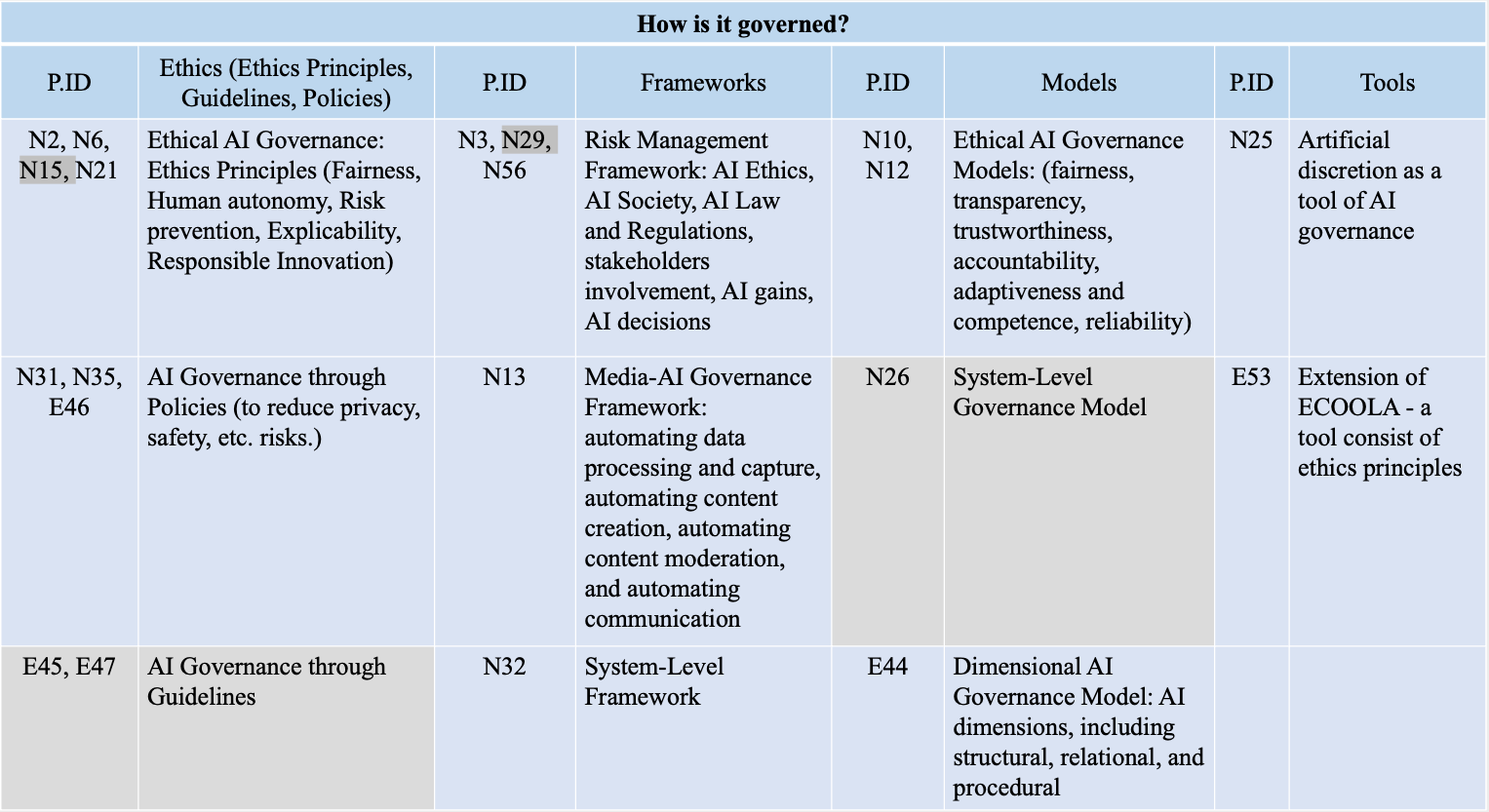}
  \caption{3W1H: \textit{How is it governed?}}
      \label{fig:how}
\end{figure*}

% \begin{figure}
%     \centering
%     \includegraphics[width=0.45\textwidth]{3W Final.png}
%      \includegraphics[width=0.45\textwidth]{1H-LAST-LAST-LAST.png}
%     \caption{3W1H Analysis (Stacked Image): Who, What, and When (orange, green, and yellow color) - How (blue color)}
%     \label{fig:six}
% \end{figure}

\subsubsection{Comprehensive analysis of 3W1H:}
Figure~\ref{fig:six} and Figure~\ref{fig:how} present a comprehensive analysis of 3W1H, applied to all 61 papers, with the paper IDs representing E as empirical studies and N as non-empirical studies from a total of 61 studies. Out of 61 studies, only five [N15], [N26], [N29], [E45], and [E47] had all of the questions in 3W1H answered; these studies are indicated in grey in figure~\ref{fig:six} and figure~\ref{fig:how}. There are a significant number of studies (20 out of 61) focused on answering the question of "how" AI should be governed (figure~\ref{fig:how}) while neglecting other aspects such as "who" should be responsible for governance and "at what stage" of the AI development life cycle. It has been observed that 11 studies (figure~\ref{fig:six}) have answered to the question "Who is governing?" where the studies [N8] and [E37] added that national and international governance coordinating committee members should be in charge of governing the AI systems; the study [N15] mentioned multiple parties should be in charge of governing data at the design, development, and deployment stages of the AI life cycle. However, this study [N15] has not specifically mentioned the names of the parties. The remaining studies (figure~\ref{fig:six}) have mentioned the multi-disciplinary steering committee ([E45]), the data management group ([N26] [N29]), and AI scientists as high level expert groups ([N2] [N3]) as stakeholders who should be responsible for governing data, processes, or systems under the AI ecosystem. It has been analysed that under the question "What is being governed?," not a single study out of 61 has mentioned the human pillar to be considered to be governed, and only one study [N21] mentioned that responsible research and innovation officers should take responsibility for governing the processes of AI development and deployment. On the other hand, 11 studies (figure~\ref{fig:six}) mentioned data pillar that should be governed, and 9 studies (figure~\ref{fig:six}) mentioned that AI systems should be governed. There are three studies common in both pillars of data and systems, i.e., [N29], [E46], and [E47]. Before the analysis of the question "How is it governed?," is presented, it has been observed that nine studies (figure \ref{fig:six}) have mentioned the different AI development stages at which governance should be done under the question "When is being governed?". For example, the study [N10] mentioned that governance of AI should be done at the initial (planning, approval, and data collection), design and deployment (monitoring and use) stages. There are two studies [N32 and E53] that mentioned that governance of AI should be done at all stages of the AI development life cycle. It is important for an organisation to know who should be responsible of governing what, and when. If it is unclear to an organisation which stakeholders they should ask to take responsibility for governing AI at various AI development stages, it is hard to reduce the risks associated with AI and ethics. If an organisation is not sure exactly what they need to govern—data, systems, processes, or humans— how can such an organisation reduce bias, establish accountability, fairness, transparency, etc.?

It has been noted that nine studies (figure~\ref{fig:how}, under Ethics) out of 61, have concentrated on developing ethical principles (fairness, privacy, safety), rules, or guidelines to assist organisations in running ethical governance for their AI systems. Three of the five ([N15], [N26], [N29], [E45], and [E47]) studies which has given answers to all questions of  Who, What, When, and How, have proposed various ethical principles and a set of guidelines for governing AI systems and data ([N15], [E45], and [E47]). Four studies [N15], [N26], [N29], and [E47] mentioned that their proposed ethical principles, and guidelines are to govern data in the AI ecosystem and one study [E45] presents a governance solution to regulate the AI system. It has been analysed that two studies [N29] and [E47] have mentioned that their proposed solution is to govern data as well as systems. This means data and systems are receiving more attention than processes and humans (figure~\ref{fig:six}). The limited focus on processes and humans in these studies suggests a potential gap in the current discourse surrounding AI governance, highlighting the need for more comprehensive consideration of ethical principles and guidelines that encompass the entire AI development life cycle, including the human dimension and associated processes. 

\subsection{AI Governance Levels}
\begin{figure}
\centering
  \includegraphics[width=0.45\textwidth]{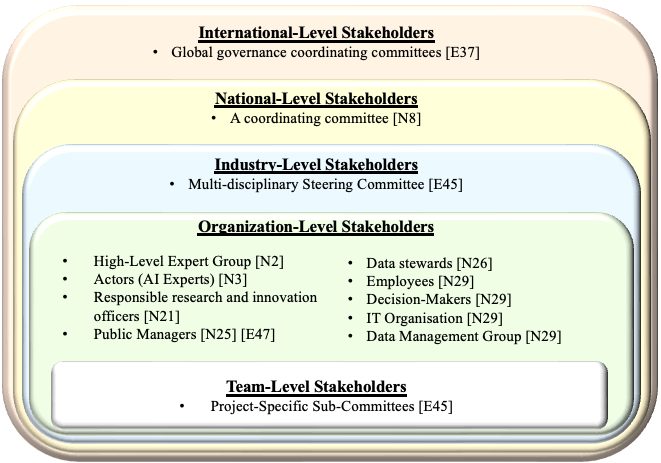}
  \caption{Stakeholder Classification}
      \label{fig:seven}
\end{figure}

\begin{figure}
\centering
  \includegraphics[width=0.4\textwidth]{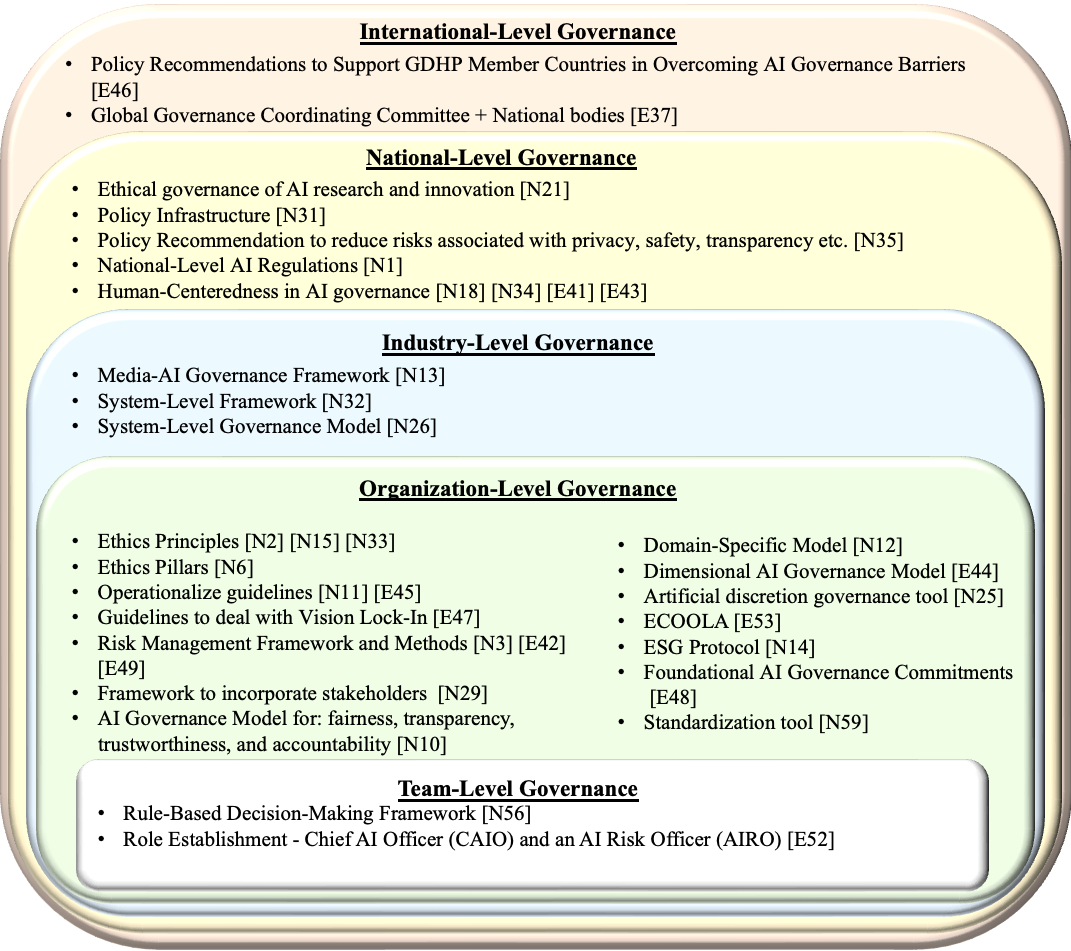}
  \caption{AI Governance Levels}
      \label{fig:eight}
\end{figure}

Inspired by the pattern catalogue proposed by Lu et al. \cite{lu2022responsible}, the stakeholders found under the question "Who is governing?" in this paper are classified under the five groups, which are shown in Figure~\ref{fig:seven}. Similarly, AI governance solutions found under the question "How is it governed?" are grouped under team-level, organisation-level, and industry-level governance. National-level and international-level governance are also added and used for further categorisation, as shown in Figure~\ref{fig:eight}. 

It has been observed that the highest number (19) (figure~\ref{fig:eight}), of AI governance solutions are categorised under organisational-level governance. Out of these 19 organisational-level governance, only 4 [N15] [N29] [E45] [E47] had answers to all questions in 3W1H, which means that they are clear on who should be responsible for regulating the AI through these solutions and at what development stage. Unfortunately, the remaining 15 studies do not offer complete solutions to AI governance to develop and deploy AI responsibly. It has also been analysed that eight AI governance solutions are categorised as national-level governance and two solutions are categorised under international-level governance. Unfortunately, none of these solutions, whether at the national or international level, have provided with complete 3W1H questions. 

It's important to recognise that, alongside organisational governance, the collaboration between national and international governance plays a pivotal role \cite{wallach2018agile}. Coordinating national bodies with international efforts enables the formulation of best AI governance solutions, ensuring responsible behaviour and ethical standards in the global AI landscape \cite{wallach2018agile}. 
% Similarly, seven studies (refer to figure three, under how is it governed - Recommendations) recommended using existing ethics rules or policies to regulate AI systems. Many researchers [N1] [N11] [N33] [E42] are focusing on ethical principles or policies to build or use existing ones, as ethical principles play a vital role in contributing to responsible AI. There is still a need for effective solutions for AI governance to mitigate the risks associated with AI. The ethical guidelines and principles that regulate AI to make it responsible should garner attention; researchers should start focusing on all facets of AI governance, such as roles that need to be involved in applying these regulations or ethics principles and which stages of the AI development lifecycle these principles are most appropriate for use. The next sub-section presents the categorization of 1W1H: Who is governing? and How is it governed? under the governance pattern for responsible AI. 

\section{Research Results}
This section presents the results and findings of each research question. \\
\textbf{RQ1: What governance frameworks, models, tools, and Policies (regulations, ethics principles, or guidelines) for AI are offered in the literature?}\\
Table~\ref{table:one} presents the frameworks for AI governance proposed in the literature along with their paper IDs (E Empirical studies, N: Non-Empirical studies). Five studies have proposed frameworks as a solution to regulate artificial intelligence systems. However, it has been observed from 3W1H analysis (Figure~\ref{fig:six} and \ref{fig:how}) that only one article [N29] out of five studies on AI governance frameworks has added answers to all questions of Who, What, When, and How. It is essential to inform organisations about who should be involved in governing AI through these frameworks and at which development stages as shown in the study by Sidorova and Saeed \cite{sidorova2022incorporating}. It has also been noted that only one study (N3) has focused on AI ethics (fairness, transparency, accountability) in its proposed framework, while the remaining four studies have proposed frameworks focusing on incorporating stakeholders in governance stages, system-level frameworks to enhance governance processes in the financial sector, and governing data in the media and communication sector, among others. It is essential to put a focus on AI ethics while proposing frameworks for governing AI to make AI responsible. If AI ethics, regulations, or responsible AI principles are ignored by researchers while proposing solutions to govern AI, then it is hard to achieve solutions that make AI ethical and responsible \cite{liao2022governance}.

\begin{table}
\centering
\begin{tabular}{|l|l|}
\hline
\multicolumn{1}{|c|}{\textbf{AI Governance Frameworks}}                                                                                                                       & \multicolumn{1}{c|}{\textbf{Paper ID}} \\ \hline
\begin{tabular}[c]{@{}l@{}}Risk Management Framework: AI Ethics, \\ 
AI Society, and AI Law and Regulations.\end{tabular}                                                                                                                  &  N3                                      \\ \hline
\begin{tabular}[c]{@{}l@{}}Media-AI Governance Framework: \\ 
automating data processing and capture, \\ 
automating content creation, automating content \\
moderation, and automating communication.\end{tabular}                                                               & N13                                    \\ \hline
\begin{tabular}[c]{@{}l@{}}Framework to incorporate stakeholders \\
along with AI risks, AI gains, and AI decisions.\end{tabular}                                                                                                     & N29                                   \\ \hline
\begin{tabular}[c]{@{}l@{}}System-Level Framework with modular building \\
blocks for efficient AI governance \\
in financial services.\end{tabular}                                                                & N32                                     \\ \hline
\begin{tabular}[c]{@{}l@{}}Fusion Fuzzy Multiple Rule-Based \\
Decision-Making Framework.\end{tabular}  & N56                   
                                                           \\ \hline
\end{tabular}
\caption{AI Governance Frameworks}
\label{table:one}
\end{table}

Table~\ref{table:two} presents the AI governance models proposed in the literature. Four studies have proposed AI governance models, and two [N10] and [N12] of the models have focused on AI ethics principles, including fairness, transparency, accountability, reliability, and so on. Only one paper [N26] of these four has answered all questions in 3W1H. The remaining three have only proposed the models without taking into consideration how important it is to know who should be responsible for regulating AI through their proposed models and at which AI development stages. 

\begin{table}[]
\centering
\begin{tabular}{|l|l|}
\hline
\multicolumn{1}{|c|}{\textbf{AI Governance Models}}                                                                                                                       & \multicolumn{1}{c|}{\textbf{Paper ID}} \\ \hline
\begin{tabular}[c]{@{}l@{}}Model consists of four different components: \\
fairness, transparency, trustworthiness, \\
and accountability.\end{tabular}                                                                                                                 &  N10                                      \\ \hline
\begin{tabular}[c]{@{}l@{}}Domain specific Model consists of three  \\
different components: adaptiveness and \\
competence, reliability and \\
transparency, and responsibility \\
and accountability.\end{tabular} 
                                                           & N12                                   \\ \hline
\begin{tabular}[c]{@{}l@{}}System-Level Governance Model for data\\
governance within BDAS - Big Data \\
Algorithmic Systems.\end{tabular}                                                                                                 & N26                                 \\ \hline
\begin{tabular}[c]{@{}l@{}}Dimensional AI Governance Model consists\\
of different AI dimensions, including \\
structural, relational, and procedural.\end{tabular} 
                                                    & E44                                 
                                                           \\ \hline
\end{tabular}
\caption{AI Governance Models}
\label{table:two}
\end{table}

It has been analysed that only two studies [N25] and [E53] have discussed and proposed AI governance tools. Study [N25] has discussed "artificial discretion,” which can be used as a tool of AI governance that helps public managers think about the implications of AI as they decide whether and how to deploy it. The second study [E53] has discussed ECOOLA, a tool that consists of ethical principles to be used to regulate AI systems. The research identified an accountability gap in ECOOLA, which was addressed through the adoption of information governance practices in ECOOLA to make its information robust. It has also been analysed that none of the studies on AI governance tools have provided all the answers to questions in 3W1H except on "how."

% \begin{table}[]
% \centering
% \begin{tabular}{|l|l|}
% \hline
% \multicolumn{1}{|c|}{\textbf{AI Governance Tools}}                                                                                                                       & \multicolumn{1}{c|}{\textbf{Paper ID}} \\ \hline
% \begin{tabular}[c]{@{}l@{}}Artificial discretion” as a tool of \\
% AI governance that helps public managers \\
% in thinking about the implications of AI as \\
% they decide whether and how to deploy it.\end{tabular}                                                                  &  N25                                      \\ \hline
% \begin{tabular}[c]{@{}l@{}}Extension of ECCOLA - a tool consist of \\
% ethics principles assist organisation in \\
% the development of ethical AI, but it has\\
% some issues of accountability which is been \\
% addressed by incorporating Information \\
% Governance practices in ECOOLA to make \\
% its information 
% robustness.\end{tabular} 

%                                                            & E53                                  \\ \hline
% \end{tabular}
% \caption{AI Governance Tools}
% \label{table:three}
% \end{table}

Ethics principles, standards, or regulations play a vital role in making AI responsible \cite{clarke2019principles}, and using ethics principles in governance practices helps contribute to responsible AI governance. Table~\ref{table:four} shows the studies (9) that have proposed different ethics principles and guidelines for the governance of AI to develop and deploy AI ethically and responsibly. There are four studies [N2], [N6], [N15], and [N21]  that have proposed different ethics principles and ethics pillars to put focus on responsible AI governance. Similarly, there are three studies [N31], [N35], and [E46] that have focused on policies that can assist organisations in mitigating the risks of privacy, safety, and transparency issues. The remaining two studies [E45] and [E47] have proposed different sets of guidelines, and one of these studies, i.e., [E45], has discussed operationalizing guidelines through different practices of value stream. It has also been analysed that three studies [N15], [E45], and [E47] of these nine have answered all questions in 3W1H. 

\begin{table}[]
% \begin{adjustbox}{width=\columnwidth}
% \resizebox{.8\textwidth}{!}{%
\begin{tabular}{|l|l|}
\hline
\multicolumn{1}{|c|}{\textbf{\begin{tabular}[c]{@{}c@{}}Policies \\ (regulations, ethics principles, or guidelines) \\ for AI Governance\end{tabular}}}                                                                                                                                                                                                                                                                                                            & \multicolumn{1}{c|}{\textbf{Paper ID}} \\ \hline
\begin{tabular}[c]{@{}l@{}}Ethics Principles proposed.\end{tabular} 
& N2                                  \\ \hline

\begin{tabular}[c]{@{}l@{}}Ethics Pillars proposed.\end{tabular}                                                                                                                                                                                                   & N6                                     \\ \hline
\begin{tabular}[c]{@{}l@{}}Ethical Values Ethical Principles: Ethical Values; \\ Ethical Principles; Guide; Ethical Norms.\end{tabular}                                                                                                                                                             & N15                                    \\ \hline
\begin{tabular}[c]{@{}l@{}}ETHNA: Ethical governance of AI research \\ and innovation.\end{tabular}                                                                                                                                                                                                                                                      & N21                                    \\ \hline
\begin{tabular}[c]{@{}l@{}}Policy Infrastructure: design, develop, implement, \\ and evaluate policies.\end{tabular}                                                                                                                                                                                                                                                                        & N31                                    \\ \hline
\begin{tabular}[c]{@{}l@{}}Policy Recommendations for privacy, safety, etc. \\ risks.\end{tabular} & N35                                    \\ \hline
\begin{tabular}[c]{@{}l@{}}Value stream practices, to operationalize the \\ guidelines.\end{tabular}                                                                                                                                                                                                                                                                                                                                             & E45                                    \\ \hline
\begin{tabular}[c]{@{}l@{}}Policy Recommendations to support GDHP \\ member countries.\end{tabular}                                                                                                                                                                                                                                       & E46                                    \\ \hline
\begin{tabular}[c]{@{}l@{}}AI Governance Guidelines proposed.\end{tabular}                                                                                                                                                                                                                                                                                                     & E47                                    \\ \hline
\end{tabular}
\caption{AI Governance Policies}
\label{table:four}
% \end{adjustbox}
\end{table}

To summarise, while frameworks, models, tools, or policies are proposed, there is a need for enhanced clarity, particularly on ethical principles and the involvement of stakeholders at different AI development stages. The prevalence of policies in the literature underscores their pivotal role in fostering responsible AI. Emphasising comprehensive details on regulatory roles and development stages is crucial for ensuring effective governance and accountability in responsible AI governance. \\

\begin{figure}
\centering
  \includegraphics[width=0.48\textwidth]{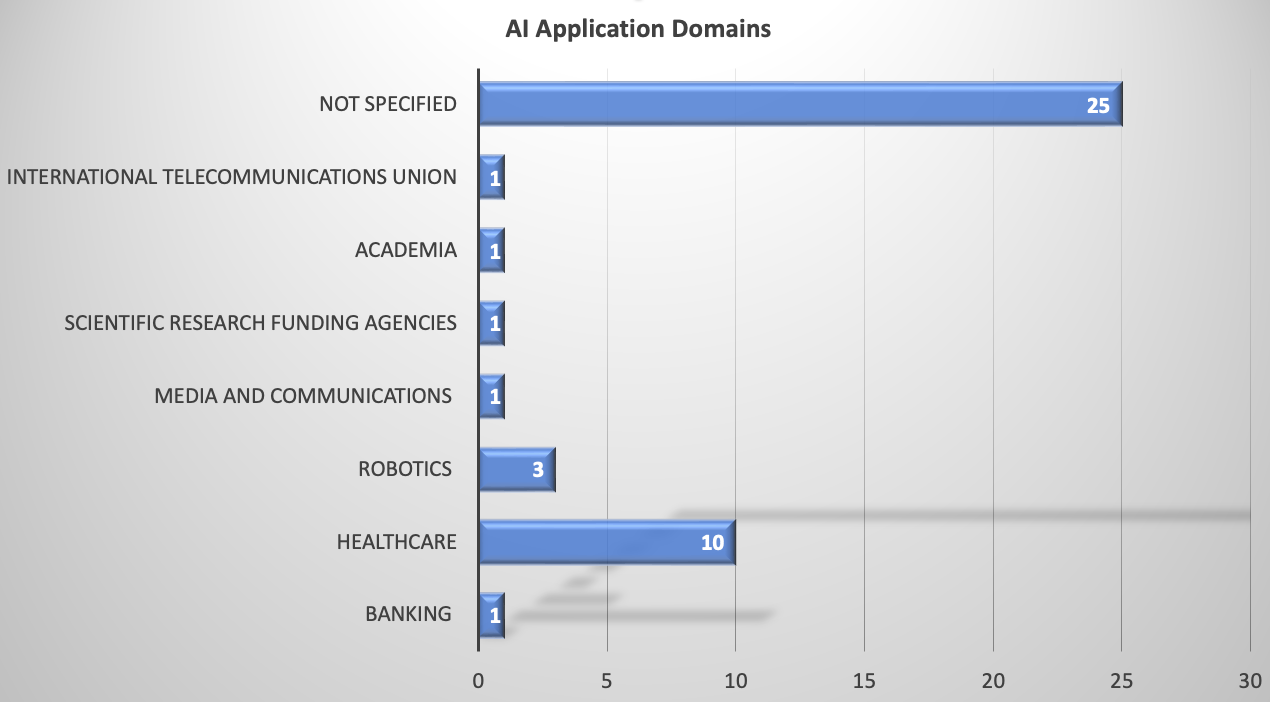}
  \caption{AI Application Domains}
      \label{fig:domains}
\end{figure}

\textbf{RQ2: Which target AI application domains are considered in the existing AI governance approaches/solutions?} \\
The AI governance frameworks, models, tools and policies discussed above have been analysed further to explore the AI application domains where these governance solutions are targeted to apply. Figure \ref{fig:domains} presents the results of the AI application domains to clearly demonstrate which domain has been targeted more and which one has been less focused. It has been analysed that the most common AI application domains targeted in the current AI governance solutions are healthcare and robotics. The results (Fig.  \ref{fig:domains}) show that 10 of the research studies on frameworks, models, tools, and policies has targeted healthcare as the application domain. Robotics is targeted in three studies. Other domains include banking, media and communication, academia, and so on. There are 25 studies that have not specified the application domains. It is necessary to do research on defining the AI application domains in order to make it obvious which application domains may utilise which AI governance frameworks, models, tools, and policies. \\

\textbf{RQ3: What are the limitations and challenges of AI governance discussed in the literature?} \\
AI governance solutions are required in order to mitigate the risks associated with AI, such as bias, discrimination, lack of transparency, and unintended harmful impacts \cite{nationalai2023} \cite{zowghi2023diversity}. But there are challenges associated with AI governance solutions as well, which are discussed in the literature and are presented in table~\ref{table:challenges}. Ethical challenges, including fairness, transparency, privacy, and trust, are the most discussed in AI governance according to the literature [N4], [N9], [E40], [E49], and [N60]. Lack of human involvement in current AI governance approaches makes the AI system less human-centric [N4] [E40] [E41]. It is important to keep human-centric concept in mind while proposing solutions to regulate AI systems as it assist in building trust in AI technology and mitigate issues of discrimination, fairness, and bias [N4] [E40] [E41].

The AI governance challenges added in table~\ref{table:challenges} are essential to consider while proposing an AI governance solution. The complex AI ethical issues covered by almost all of the studies have yet to be resolved through highly appropriate AI governance practises. The current AI governance frameworks, models, tools, and policies do cover the risk layer to mitigate the different risks associated with the application of artificial intelligence in different domains and make AI responsible \cite{sidorova2022incorporating}, \cite{wirtz2020dark},  \cite{unver2023governing}. However, there are still challenges in completely addressing these ethical concerns as shown in table~\ref{table:challenges}. The next section concludes the paper with highlights of future work.

\begin{table}[]
\begin{tabular}{|l|l|}
\hline
\multicolumn{1}{|c|}{\textbf{Challenges of AI Governance}}                                             & \multicolumn{1}{c|}{\textbf{Paper ID}}                                        \\ \hline
\begin{tabular}[c]{@{}l@{}}Ethical challenges (fairness, transparency, \\ privacy, trust)\end{tabular} & \begin{tabular}[c]{@{}l@{}}N4, N9 \\ E40, E49\\ N60\end{tabular} \\ \hline
\begin{tabular}[c]{@{}l@{}}Legal-regulatory challenges (human-centric \\ concerns)\end{tabular}        & \begin{tabular}[c]{@{}l@{}}N4, E40\\ E41\end{tabular}                  \\ \hline
Technical challenges (biases and harms)                                                                & N4                                                                         \\ \hline
Societal and Economic Challenges                                                                       & N5, E40                                                                  \\ \hline
\end{tabular}
\caption{Challenges of AI Governance}
\label{table:challenges}
\end{table}

\section{Conclusion and Future Work}
In conclusion, this systematic literature review of 61 selected studies achieves its objectives by providing a comprehensive summary of current AI governance solutions, including frameworks, tools, models, and policies. The challenges within existing AI governance solutions are also presented in this paper, emphasizing the prevalent focus on ethical considerations such as fairness, transparency, privacy, and trust. The 3W1H analysis reveals a critical gap in addressing "who," "when," and "what" aspects of AI governance in a holistic manner, with a disproportionate emphasis on the "how." Only 5 studies of 61 have provided all the answers to questions in 3W1H. The analysis has also revealed that the highest number of governance solutions are under organisational-level governance. While ethical principles play a vital role, there is a call for enhanced clarity, especially in stakeholder involvement at different AI development stages. The findings emphasize the imperative for AI governance solutions to address not only risks but also crucial ethical aspects, promoting responsible and effective governance in the dynamic landscape of artificial intelligence. This underscores the need for future AI governance solutions to align closely with RAI principles, fostering a holistic and principled approach to the governance of artificial intelligence.

As part of future work, an extended analysis is delving into grey literature, including white papers, reports, and company AI governance frameworks. This exploration aims to extract additional insights, contributing to a more nuanced understanding of responsible AI governance and enriching the discourse on AI governance practices, their implications, and ethical considerations.

\bibliographystyle{unsrt}
\bibliography{reference.bib}

\appendix
\label{appendix}  
\end{document}